\documentclass[conference]{IEEEtran}
\IEEEoverridecommandlockouts
\usepackage{amsmath,amssymb,amsfonts}
\usepackage{algorithm}
\usepackage{balance}
\usepackage{hyperref}
\usepackage{algpseudocode}
\usepackage{graphicx}
\newcommand\blfootnote[1]{
    \begingroup
    \renewcommand\thefootnote{}\footnote{#1}
    \addtocounter{footnote}{-1}
    \endgroup
}
\usepackage{textcomp}
\usepackage{xcolor}
\usepackage{array}
\usepackage{subcaption}
\usepackage{booktabs}
\usepackage{tikz}
\usetikzlibrary{arrows.meta, positioning, fit, backgrounds, calc, matrix}

\def\BibTeX{{\rm B\kern-.05em{\sc i\kern-.025em b}\kern-.08em
    T\kern-.1667em\lower.7ex\hbox{E}\kern-.125emX}}
\begin{document}

\title{Decoupled I/O-Dominant Pipelines for Large-Scale Whole-Slide Image Embedding Extraction
}

\author{\IEEEauthorblockN{Mayanka Chandrashekar}
\IEEEauthorblockA{\textit{Oak Ridge National Laboratory} }
\and
\IEEEauthorblockN{Xi Zhang}
\IEEEauthorblockA{\textit{Oak Ridge National Laboratory} }
\and
\IEEEauthorblockN{Ethan Seefried}
\IEEEauthorblockA{\textit{Oak Ridge National Laboratory} }
\and
\IEEEauthorblockN{\qquad \qquad  Tirthankar Ghosal}
\IEEEauthorblockA{\qquad \qquad  \textit{Oak Ridge National Laboratory} }
\and
\IEEEauthorblockN{John Gounley}
\IEEEauthorblockA{\textit{Oak Ridge National Laboratory} }
\and
\IEEEauthorblockN{Heidi Hanson}
\IEEEauthorblockA{\textit{Oak Ridge National Laboratory} }
}

\maketitle

\begin{abstract}




Whole-slide images (WSIs) are central to computational pathology but are prohibitively large, making patch-based processing the practical unit for foundation model inference. At scale, however, generating and handling massive numbers of patches on quickly introduces significant I/O and orchestration overhead, often dominating end-to-end performance.
We present a decoupled, I/O-aware pipeline for large-scale WSI embedding extraction that decomposes the workflow into three stages: (1) patch generation and staging, (2) embarrassingly parallel embedding inference, and (3) sharded vector database ingestion. This design isolates data movement from compute, enabling efficient patch delivery, scalable multi-node inference with minimal communication.
The resulting system produces a distributed vector database where embeddings are persistently coupled with rich metadata (e.g., patient, slide, and patch attributes), enabling efficient filtering, retrieval, and downstream reuse. This representation database is compact and reusable for tasks such as retrieval, classification, and few-shot learning, particularly benefiting low-resource environments.
We show that decoupling I/O, computation, and ingestion enables high-throughput WSI embedding extraction at scale. By characterizing the scaling envelope, we demonstrate that storage dominates beyond moderate concurrency, reframing WSI embedding extraction as a data-centric systems problem rather than a purely compute-bound workload.

\end{abstract}

\blfootnote{Notice: Office of Science of the U.S. Department of Energy. This manuscript has been authored by UT-Battelle, LLC, under contract DE-AC05-00OR22725 with the US Department of Energy (DOE). The US government retains and the publisher, by accepting the article for publication, acknowledges that the US government retains a nonexclusive, paid-up, irrevocable, worldwide license to publish or reproduce the published form of this manuscript, or allow others to do so, for US government purposes. DOE will provide public access to these results of federally sponsored research in accordance with the DOE Public Access Plan (http://energy.gov/downloads/doe-public-access-plan). }


\section{Introduction}
Whole-slide images (WSIs) are a cornerstone of computational pathology, enabling high-resolution analysis of tissue at gigapixel scale. The emergence of foundation models has further increased the demand for large-scale embedding extraction from WSIs, supporting downstream tasks such as retrieval, classification, and cohort-level analysis. However, the sheer size of WSIs—often exceeding tens of gigapixels per image—makes direct processing infeasible, necessitating patch-based decomposition as the fundamental unit of computation.

While patching enables scalable model inference, it also fundamentally reshapes the systems problem. At moderate scale, patches can be generated and processed on demand with acceptable overhead. At large scale, the need to generate, move, and process millions of patches introduces significant data orchestration challenges. In particular, patch generation from large WSIs, coordination of data movement across storage and compute nodes, and efficient handling of intermediate representations can dominate runtime, shifting the bottleneck away from accelerator throughput toward I/O and pipeline design.

Existing WSI pipelines are typically designed as tightly coupled workflows in which patch generation, model inference, and output handling are interleaved.  While such designs simplify implementation, they often lead to inefficient resource utilization, increased contention on shared storage systems, and limited scalability. Furthermore, these pipelines frequently treat embedding outputs as transient artifacts, missing the opportunity to store reusable embeddings that can be used for downstream applications and reduce the cost of future large-scale computation.

We define an I/O-dominated workload as an end-to-end runtime governed by data movement and storage system behavior rather than compute throughput or communication overhead. By data-centric, we refer to workloads where performance is determined primarily by data orchestration—generation, movement, and persistence—rather than computation or communication. Large-scale WSI embedding extraction is an I/O-dominated, data-centric workload in which performance is governed by data movement and storage system behavior rather than compute or communication.

In this work, we show the utility of treating large-scale WSI embedding extraction as a data-centric systems problem, rather than a purely compute-driven workload. To this end, we present a decoupled pipeline architecture that separates the workflow into three stages: (1) patch generation and I/O-aware data staging, (2) massively parallel embedding inference, and (3) sharded vector database ingestion. This decomposition enables each stage to be optimized independently according to its dominant constraints—storage throughput, compute parallelism, and data management—while minimizing cross-stage interference.

A central aspect of our design is the treatment of the vector database as a reusable resource to be shared with the scientific and medical communities. Rather than producing embeddings as temporary files used for a specific project, our pipeline directly ingests them into a distributed, sharded vector database, with rich metadata (e.g., patient, slide, and patch-level attributes) persistently associated with each embedding. This creates a shared resource that supports efficient indexing, filtering, and retrieval, converting the output of large-scale computation into a reusable, queryable, and energy-efficient resource that accelerates downstream analyses. A stored embedding resource is also particularly valuable in low-resource environments, where extracting embeddings from a large number of WSIs is impractical.

We implemented our pipeline on Oak Ridge Leadership Computing - Frontier Exascale System \cite{Frontier23} and evaluated it on a large-scale WSI dataset. Our results show that decoupling patch orchestration, inference, and ingestion enables high throughput, scalability, and resource efficiency. More broadly, our work demonstrates that aligning pipeline structure with data movement characteristics is necessary for scaling data-intensive AI workloads.

This paper makes the following contributions:

\begin{itemize}
    \item We identify and characterize WSI embedding extraction as an I/O- and orchestration-dominated workload at large scale. 
    \item We design and implement a three-stage decoupled pipeline that separates patch generation, parallel inference, and sharded vector database ingestion.
    \item We introduce the construction of metadata-aware vector databases as a reusable resource for large-scale WSI processing pipelines.
    \item We demonstrate improved scalable execution and characterize resource bottlenecks on supercomputing infrastructure.
\end{itemize}

\section{Background, Motivation and Related Work}
\begin{table*}[t]
\centering
\footnotesize
\renewcommand{\arraystretch}{0.8}
\setlength{\tabcolsep}{3.5pt}
\begin{tabular}{p{2.2cm}|p{2.2cm}|p{1.8cm}|p{1.7cm}|p{2.4cm}|>{\centering\arraybackslash}p{1.15cm}|>{\centering\arraybackslash}p{1.05cm}|>{\centering\arraybackslash}p{1.2cm}|>{\centering\arraybackslash}p{1.5cm}}
\toprule
\textbf{Category} & \textbf{Model / System} & \textbf{Learning Type} & \textbf{Data Scale} & \textbf{Key Strength} & \textbf{Pipeline Coupling} & \textbf{I/O Awareness} & \textbf{Scalability} & \textbf{Embedding Reuse} \\
\midrule

\multicolumn{9}{c}{\textit{WSI Representation Models Used in This Work}} \\
\midrule

Hierarchical
& HIPT \cite{HIPT22}
& Hierarchical SSL
& 10,678 WSIs
& Multi-scale tissue modeling
& N/A
& $\times$
& N/A
& Partial \\

Foundation Model
& Virchow / Virchow2 \cite{Virchow24}
& Large-scale SSL
& 1.5M--3.1M+ H\&E slides
& Strong general-purpose features
& N/A
& $\times$
& N/A
& Partial \\

Foundation Model
& H-Optimus / H-Optimus-0 \cite{hoptimus0}
& Pathology FM
& Large-scale histopathology corpus
& Pathology-specialized representation learning
& N/A
& $\times$
& N/A
& Partial \\

\midrule
\multicolumn{9}{c}{\textit{WSI and HPC Pipelines}} \\
\midrule

Traditional Pipelines
& OpenSlide-based workflows \cite{OpenSlide13}
& Patch extraction + inference
& Variable
& Simple and widely used
& Tight
& $\times$
& Limited
& $\times$ \\

HPC Pipelines
& GEMS \cite{GEMS20}, scalable WSI HPC pipeline \cite{9835363}
& Distributed training / preprocessing
& Large-scale WSI workloads
& Demonstrates HPC viability for WSI-scale data
& Partial
& Partial
& High
& $\times$ \\

\midrule
\multicolumn{9}{c}{\textit{HPC and Data Systems}} \\
\midrule

HPC Workflows
& Pegasus \cite{Pegasus15}
& Workflow DAGs
& Large-scale science
& Decoupled execution
& Decoupled
& \checkmark
& High
& N/A \\

I/O-Aware Systems
& Burst buffers, I/O studies \cite{shan08,6232369}
& Storage optimization
& HPC workloads
& Data movement optimization
& Decoupled
& \checkmark
& High
& N/A \\

Vector Databases
& FAISS / Milvus \cite{FAISS19,2021milvus}
& ANN indexing
& Billion-scale vectors
& Fast retrieval
& Separate
& \checkmark
& High
& \checkmark \\

WSI Tooling
& LazySlide \cite{zheng2026lazyslide}
& Interoperable WSI software
& Large-scale WSI analysis
& Integration with downstream multimodal workflows
& Partial
& Partial
& Moderate
& \checkmark \\

\midrule
\multicolumn{9}{c}{\textbf{This Work}} \\
\midrule

Decoupled WSI Pipeline
& Ours
& Multi-stage pipeline
& 100M+ patches
& I/O-dominant scaling with reusable embeddings
& \textbf{Fully decoupled}
& \textbf{\checkmark}
& \textbf{High (embarrassingly parallel)}
& \textbf{\checkmark\ (persistent + metadata)} \\

\bottomrule
\end{tabular}
\caption{Comparison of representative WSI models, WSI/HPC pipelines, and data systems. Modern pathology models improve representation quality but generally assume efficient data access. Our work targets the orthogonal systems problem: scalable, I/O-aware embedding extraction with persistent reuse.}
\label{tab:comparison}
\end{table*}

Recent advances in computational pathology have been driven by large-scale representation learning for WSIs. Early approaches relied on weak supervision and multiple-instance learning to aggregate patch-level features into slide-level predictions \cite{Campanella19, Lu21}. 

Subsequent work introduced self-supervised learning and vision transformer-based approaches that improved representation quality and generalizability \cite{DINOv2}. Hierarchical architectures such as HIPT further model multi-scale structure, capturing both cellular and tissue-level context \cite{HIPT22}.
More recently, whole-slide foundation models have emerged as the dominant paradigm. Models such as Virchow and H-Optimus-0 leverage large-scale pretraining on histopathology datasets to achieve strong performance across tasks \cite{Virchow24, hoptimus0}. Multimodal extensions (e.g., TITAN, mSTAR) integrate WSIs with clinical reports and molecular data, enabling retrieval, pathology report generation, and cross-modal reasoning \cite{TITAN,mSTAR}. Despite these advances, prior work largely focuses on improving model accuracy and do not address computational bottlenecks that slow large scale deployment.

On the systems side, WSI pipelines are typically implemented as tightly coupled workflows in which patch generation, model inference, and output handling are interleaved. While such designs simplify implementation, they limit scalability due to shared I/O contention and prevent independent optimization of pipeline stages. As noted in our pipeline formulation, WSI processing naturally decomposes into patch generation, embedding inference, and data management stages with distinct resource characteristics .

In parallel, the HPC community has extensively studied data movement bottlenecks in large-scale workflows. Prior work highlights that I/O and data orchestration increasingly dominate runtime and proposes techniques such as burst buffers, hierarchical storage, and workflow decoupling \cite{shan08, 6232369, Pegasus15}. However, these approaches are designed for workloads that involve large contiguous datasets and structured communication patterns. In contrast, WSI pipelines generate massive numbers of small, independent data objects and exhibit embarrassingly parallel computation with minimal communication, creating a mismatch with existing HPC assumptions.

Recent HPC efforts have begun to explore WSI-scale workloads. Systems such as GEMS demonstrate distributed training for high-resolution pathology images \cite{GEMS20}, and subsequent work investigates scalable pipelines for gigapixel WSI processing on leadership-class systems \cite{9835363}. However, these approaches primarily focus on training or preprocessing and do not address embedding extraction pipelines or their I/O-dominant behavior.

On the data systems side, vector databases such as FAISS and Milvus enable scalable indexing and retrieval over large embedding collections \cite{FAISS19, 2021milvus}. Emerging WSI ecosystems such as LazySlide emphasize interoperability and integration with downstream multimodal workflows \cite{zheng2026lazyslide}. However, these systems remain decoupled from upstream embedding pipelines, and embeddings are typically treated as transient outputs rather than reusable resources.

What is missing is a system that treats WSI embedding extraction itself as an I/O-dominant pipeline, rather than assuming patch access is cheap. WSI models focus on representation learning, HPC systems address data movement under different assumptions, and vector databases operate independently of data generation pipelines. No prior work jointly addresses I/O-dominant embedding extraction, pipeline decoupling, and embedding reuse in large-scale WSI processing.

\section{WSI Pipeline and Parallel Execution Model}
\begin{figure*}
    \centering
    \includegraphics[width=0.9\linewidth]{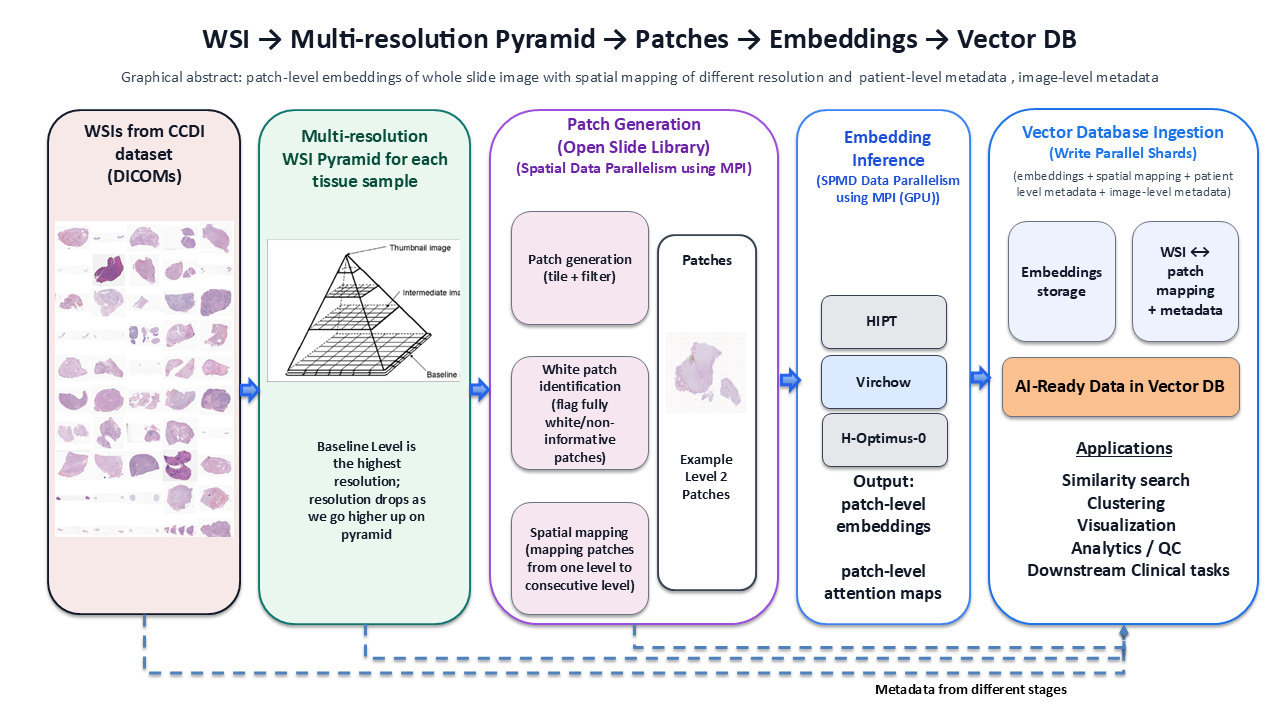}
    \caption{Decoupled WSI embedding pipeline. The workflow is organized into three stages: (1) MPI-based patch generation and staging, (2) embarrassingly parallel embedding inference across GPUs, and (3) shard-parallel vector database ingestion. Each stage is parallelized according to its dominant constraint—storage, compute, and write throughput—enabling independent scaling. The pipeline produces a persistent, metadata-aware embedding database for downstream reuse.}
    \label{fig:workflow}
\end{figure*}
Our pipeline (Figure~\ref{fig:workflow}) is organized as three decoupled stages: patch generation, embedding inference, and vector-database ingestion. The key systems design choice is that each stage is parallelized according to its dominant bottleneck rather than forcing a single end-to-end execution strategy. In the current implementation, Stage 1 uses MPI-based spatial decomposition over patch coordinates, Stage 2 uses SPMD (Single Program Multiple Data) rank-parallel inference with one data shard per task/GPU, and Stage 3 uses shard-parallel database construction. This decoupling is central to making the workflow HPC-ready because it isolates I/O-heavy expansion, accelerator-heavy inference, and write-heavy persistence into independently scalable components.

\noindent \textbf{Stage 1: Patch Generation (MPI Spatial Decomposition)}

Patch generation is implemented as an MPI program over a single WSI at a specified pyramid level. All ranks independently construct the same global grid of valid patch coordinates. Work is distributed using \emph{deterministic cyclic partitioning}, where rank $r$ processes indices $r, r+R, r+2R, \dots$, with $R$ denoting the total number of MPI ranks.

For each assigned coordinate $(x_L, y_L)$ at level $L$, the rank maps to level-0 coordinates using the slide downsampling factor, reads the corresponding patch via \texttt{read\_region}, computes a white-background fraction, and writes the patch to a rank-local directory. Metadata including spatial coordinates, centroid, and patch path are stored locally.

No communication occurs in the steady-state extraction loop. Synchronization is limited to a single barrier, after which rank 0 gathers metadata from all ranks and writes a global \texttt{index.csv}. This design avoids communication in the critical path, but introduces significant filesystem pressure due to large-scale small-file writes.

This stage exhibits \textbf{spatial data parallelism} over the WSI coordinate space. The cyclic partitioning ensures balanced work distribution even under irregular tissue layouts, while avoiding synchronization during extraction. However, performance is dominated by small-file I/O and filesystem contention rather than compute.

\begin{algorithm}[h]
\caption{Distributed Patch Generation with Cyclic Coordinate Partitioning}
\begin{algorithmic}[1]
\Require WSI $S$, level $L$, patch size $p$, stride $s$, ranks $0,\dots,R-1$
\State Each rank opens $S$
\State $\mathcal{G} \leftarrow \texttt{grid\_coords}(S, L, p, s)$
\State $\mathcal{G}_r \leftarrow \{ \mathcal{G}[i] \mid i \bmod R = r \}$
\ForAll{$(i,j,x_L,y_L) \in \mathcal{G}_r$}
    \State $d \leftarrow$ level downsample factor
    \State $(x_0, y_0) \leftarrow (\mathrm{round}(x_L \cdot d), \mathrm{round}(y_L \cdot d))$
    \State $P \leftarrow \texttt{read\_region}(S, (x_0,y_0), L, (p,p))$
    \State $\omega \leftarrow$ white fraction of $P$
    \State Save $P$ to rank-local storage
    \State Record metadata $(x_0,y_0,x_L,y_L,\omega,\texttt{path})$
\EndFor
\State MPI Barrier
\State Rank 0 gathers metadata and writes \texttt{index.csv}
\end{algorithmic}
\end{algorithm}

\noindent\textbf{Stage 2: Embedding Inference (SPMD Data Parallelism)}
Embedding extraction is executed in Single Program, Multiple Data (SPMD) fashion across distributed tasks, typically with one task per GPU. Each task independently loads the same patch index and constructs an identical dataset abstraction. Work is partitioned using \emph{strided indexing}, where task $r$ processes dataset entries $r, r+R, r+2R, \dots$.

Each rank processes its subset using a local \texttt{DataLoader}, performing batched inference with a foundation model (e.g., HIPT, Virchow, H-Optimus-0). Depending on model capabilities, embeddings and attention tensors are extracted either in a single forward pass or in separate passes.

No collective communication (e.g., all-reduce) is used during inference. Instead, each rank accumulates results locally and serializes them to disk. Synchronization is deferred to a final consolidation step, where rank 0 gathers all rank outputs via file-based coordination, merges them, sorts patches by spatial coordinates, and writes per-WSI embedding files.

This stage exhibits \textbf{embarrassingly parallel data parallelism}. Each rank operates independently with no communication in the critical path. Unlike traditional distributed deep learning, there is no parameter synchronization or gradient exchange. The only coordination occurs at the end of execution, where outputs are merged via file-based aggregation.

\begin{algorithm}[h]
\caption{SPMD Embedding Extraction with Rank-Local Shards}
\begin{algorithmic}[1]
\Require Patch index $\mathcal{I}$, model $f_\theta$, ranks $0,\dots,R-1$
\State Each rank loads $\mathcal{I}$ and constructs dataset $\mathcal{D}$
\State $\mathcal{D}_r \leftarrow \{ \mathcal{D}[i] \mid i \bmod R = r \}$
\State Initialize DataLoader over $\mathcal{D}_r$
\ForAll{mini-batch $B \subset \mathcal{D}_r$}
    \State Load patches and coordinates
    \State Apply preprocessing
    \If{model supports joint extraction}
        \State $(E, A) \leftarrow f_\theta(B)$
    \Else
        \State $E \leftarrow f_\theta^{\text{embed}}(B)$
        \If{attention supported}
            \State $A \leftarrow f_\theta^{\text{attn}}(B)$
        \EndIf
    \EndIf
    \State Append $(E, A, \text{coords})$ to local buffers
\EndFor
\State Serialize local results to rank-specific file
\State Write completion marker
\State Rank 0 loads all rank outputs, merges and sorts by coordinates
\State Write final embedding (and attention) files
\end{algorithmic}
\end{algorithm}

\noindent\textbf{Stage 3: Vector Database Ingestion (Write-Parallel Sharding)}

The final stage aggregates embeddings into a persistent vector database, enabling downstream retrieval and analytics. This stage is constrained primarily by write throughput and indexing overhead rather than compute. The shards are partitioned at file granularity to reduce coordination overhead. Vector database ingestion is implemented as streaming ingestion with batched inserts. We also have 
file-level checkpointing for fault tolerance. 

This design exposes shard-level parallelism and can scale with the number of shards until limited by filesystem throughput while preserving spatial and metadata relationships. The embedding and metadata in vector database becomes a persistent resource rather than transient outputs, allowing downstream tasks to operate directly on stored representations without repeatedly accessing raw WSIs.

\begin{algorithm}[h]
\caption{Sharded Vector Database Ingestion}
\begin{algorithmic}[1]
\Require Embedding File set $\mathcal{E}$, metadata $\mathcal{M}$, shards $K$
\ForAll{rank $r \in \{1,\dots,K\}$ \textbf{in parallel}}
    \State Initialize shard database $D_r$
    \State Assign subset $(\mathcal{E}_r, \mathcal{M}_r)$
    \ForAll{$(e_i, m_i) \in (\mathcal{E}_r, \mathcal{M}_r)$}
        \State Insert $(e_i, m_i)$ into $D_r$
    \EndFor
    \State Build local index on $D_r$
\EndFor
\State Parallel query across shards at application level or individual shard can also be queried. 
\end{algorithmic}
\end{algorithm}

\section{Experimental Setup}

\subsection{Dataset}

We evaluate the proposed pipeline on 4,185 hematoxylin and eosin (H\&E) whole-slide images from the Childhood Cancer Data Initiative Molecular Characterization Initiative, obtained via the National Cancer Institute Imaging Data Commons \href{https://ccdi.cancer.gov/MCI} ({CCDI MCI}). The dataset spans a wide range of slide sizes and tissue content, yielding multi-resolution workloads from tens to more than $10^5$ patches per WSI. This diversity enables evaluation across both small- and large-scale regimes and captures variability representative of real-world pathology cohorts.

\begin{table}[h]
    \centering
    \begin{tabular}{|c|c|} \hline
    \multicolumn{2}{|l|}{CCDI Dataset} \\ \hline
       No. of Tissue Samples  & 4185 \\ \hline
        No. of Patients & 4054 \\ \hline
        No. of DICOMS & 19,603 \\ \hline
        No. of Patches (256x256) & 414M \\ \hline
        No. of Non-White Patches & 170M \\ \hline
    \end{tabular}
    \caption{CCDI Dataset Statistics}
\end{table}

\subsection{Models}

We consider three representative foundation models—HIPT, H-Optimus-0, and Virchow2—which differ in patch resolution, embedding dimensionality, and computational characteristics. This variation allows us to evaluate whether observed system behavior is consistent across models with different compute intensity and output sizes.

\subsection{System and Hardware}

All experiments are conducted on the \textit{Frontier} system at Oak Ridge National Laboratory. Each compute node consists of a single 64-core AMD EPYC CPU paired with 8 AMD Instinct MI250X GPUs, interconnected via the Slingshot high-speed network.

The system is backed by the \textit{Orion} parallel filesystem, a Lustre-based storage system designed for exascale workloads. Orion provides high aggregate bandwidth but is subject to metadata and small-file access constraints under high concurrency. These characteristics are particularly relevant for our workload, which generates and accesses millions of small patch files concurrently across nodes.

This hardware configuration directly influences observed performance behavior: while GPU compute capacity is abundant, effective throughput is governed by filesystem bandwidth and metadata scalability. As a result, the experiments capture realistic system-level bottlenecks encountered in data-intensive workloads on leadership-class HPC systems.

\subsection{Experimental Methodology}

The evaluation covers all three stages of the pipeline. Patch generation is executed using MPI-based CPU parallelism over spatial coordinates, producing large collections of small files and exposing filesystem-level scaling behavior. Embedding extraction is evaluated under both multi-node distributed execution (for large workloads) and single-GPU per-WSI execution (for small workloads), enabling comparison across parallelization regimes. The vector database ingestion stage is evaluated using shard-parallel insertion into a distributed vector database, capturing write-intensive behavior at scale.

To isolate system bottlenecks in the embedding stage, we organize experiments into four setups. Setup A varies patch count for a fixed WSI and hardware configuration to characterize the transition from compute-bound to I/O-dominated execution. Setup B performs strong-scaling analysis by increasing GPU count for a fixed workload, revealing the impact of storage contention on parallel efficiency. Setup C uses kernel-level profiling to verify that GPU execution remains efficient and that end-to-end limitations arise from data movement rather than compute. Setup D evaluates production-scale behavior across thousands of WSIs, capturing variability in workload size, runtime, and scheduling dynamics.

Across all stages, we measure runtime, throughput, speedup, and efficiency to distinguish compute scaling from storage-limited behavior. Stage-specific metrics include file counts, output size, average file size, and write throughput for patch generation; GPU utilization, I/O fraction, and per-WSI runtime for embedding extraction; and insertion rate, total runtime, and memory footprint for vector database ingestion.

\section{Results}
We evaluate the proposed pipeline to understand how performance evolves with scale and to identify the dominant bottlenecks at each stages. Across all experiments, a consistent pattern emerges: although each stage is embarrassingly parallel in isolation, end-to-end performance is ultimately limited by data movement and storage system behavior rather than compute or communication.

\subsection{Patch Generation Scaling}

Table II shows patch-generation performance on a single node (56 MPI ranks). Increasing patch size from 224 to 256 reduces file count while keeping total output size nearly constant, resulting in consistently lower runtime and higher write throughput. For example, reducing file count by ~25\% yields proportional throughput improvements, despite a negligible change in total bytes written. This shows that performance is driven by the number of files (metadata overhead), not data volume .

Strong scaling results shown in Table III further support the conclusions that performance is bound by the number of files. Increasing ranks from 56 to 224 yields a 1.69× speedup (42.3\% efficiency), while aggregate throughput increases sublinearly. Despite the absence of communication, scaling degrades due to shared filesystem contention. These results show that patch generation transitions from communication-free overhead to I/O-dominated performance as parallelism increases, with additional resources increasing storage pressure rather than improving performance.

\begin{table*}[t]
\centering
\small
\setlength{\tabcolsep}{3pt}
\begin{tabular}{lp{1cm}ccccccp{2cm}}
\toprule
\textbf{Slide} & \textbf{Patch Size} & \textbf{Nodes} & \textbf{MPI Ranks} & \textbf{Runtime (s)} & \textbf{Files Created} & \textbf{Output Size (GB)} & \textbf{Avg File Size (KB)} & \textbf{Write Throughput (MB/s)} \\
\midrule
\texttt{ff93fcc3} & 224 & 1 & 56 & 19.82 & 93,696  & 2.42 & 25.8 & 116.22 \\
\texttt{ff93fcc3} & 256 & 1 & 56 & 17.21 & 71,773  & 2.40 & 33.4 & 133.03 \\
\texttt{d47c2741} & 224 & 1 & 56 & 40.67 & 184,785 & 6.75 & 36.5 & 158.20 \\
\texttt{d47c2741} & 256 & 1 & 56 & 34.11 & 141,561 & 6.71 & 47.4 & 187.60 \\
\texttt{ca6e4e7a} & 224 & 1 & 56 & 47.24 & 224,133 & 9.94 & 44.4 & 200.67 \\
\texttt{ca6e4e7a} & 256 & 1 & 56 & 42.02 & 171,702 & 9.88 & 57.5 & 224.27 \\
\bottomrule
\end{tabular}
\vspace{1mm}
\caption{
Stage 1 patch-generation performance for three representative slides at 56 MPI ranks using patch sizes 224 and 256.
The selected slides span smaller, medium, and larger output footprints in the benchmark set.
Increasing patch size reduces the number of generated files while increasing average file size and effective write throughput.
}
\label{tab:patch_generation_main}
\end{table*}

\begin{table*}[t]
\centering
\small
\setlength{\tabcolsep}{5pt}
\begin{tabular}{cccccccc}
\toprule
\textbf{Patch Size} & \textbf{Nodes} & \textbf{MPI Ranks} & \textbf{Runtime (s)} & \textbf{Speedup} & \textbf{Efficiency (\%)} & \textbf{Files Created} & \textbf{Write Throughput (MB/s)} \\
\midrule
224 & 1 & 56  & 22.39 & 1.00$\times$ & 100.0 & 93,696 & 102.66 \\
224 & 2 & 112 & 18.85 & 1.19$\times$ & 59.4  & 93,696 & 122.99 \\
224 & 4 & 224 & 13.22 & 1.69$\times$ & 42.3  & 93,696 & 177.50 \\
256 & 1 & 56  & 16.89 & 1.00$\times$ & 100.0 & 71,773 & 135.46 \\
256 & 2 & 112 & 13.41 & 1.26$\times$ & 63.0  & 71,773 & 171.57 \\
256 & 4 & 224 & 9.56  & 1.77$\times$ & 44.2  & 71,773 & 244.02 \\
\bottomrule
\end{tabular}
\vspace{1mm}
\caption{
Stage 1 strong-scaling results for one representative slide (ID \texttt{ff93fcc3-7bc2-4656-8e2d-e379180fa042}) using patch sizes 224 and 256 across 56, 112, and 224 MPI ranks.
Speedup and parallel efficiency are computed relative to the 56-rank baseline for each patch size.
The sublinear scaling and rising write throughput are consistent with the I/O-dominated behavior
}
\label{tab:patch_scaling}
\end{table*}

\subsection{Embedding Inference Analysis}

\textbf{Setup A: Patch Sweep — Identifying I/O-Dominated Behavior}

\begin{figure*}[t]
    \centering
    \begin{subfigure}[t]{0.45\textwidth}
        \centering
        \includegraphics[width=\linewidth]{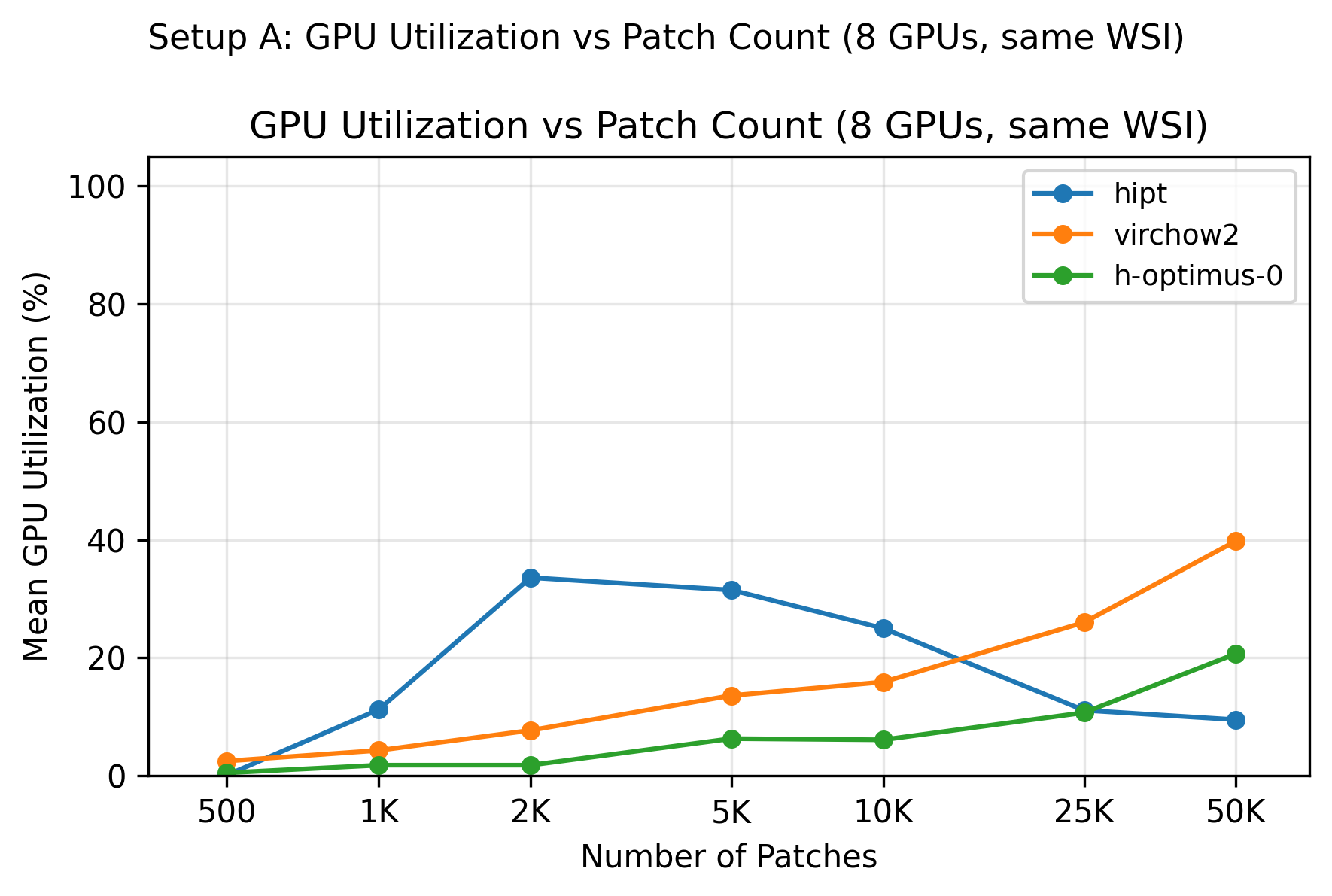}
        \caption{Mean GPU utilization.}
        \label{fig:setupA_core_gpu}
    \end{subfigure}
    \begin{subfigure}[t]{0.45\textwidth}
        \centering
        \includegraphics[width=\linewidth]{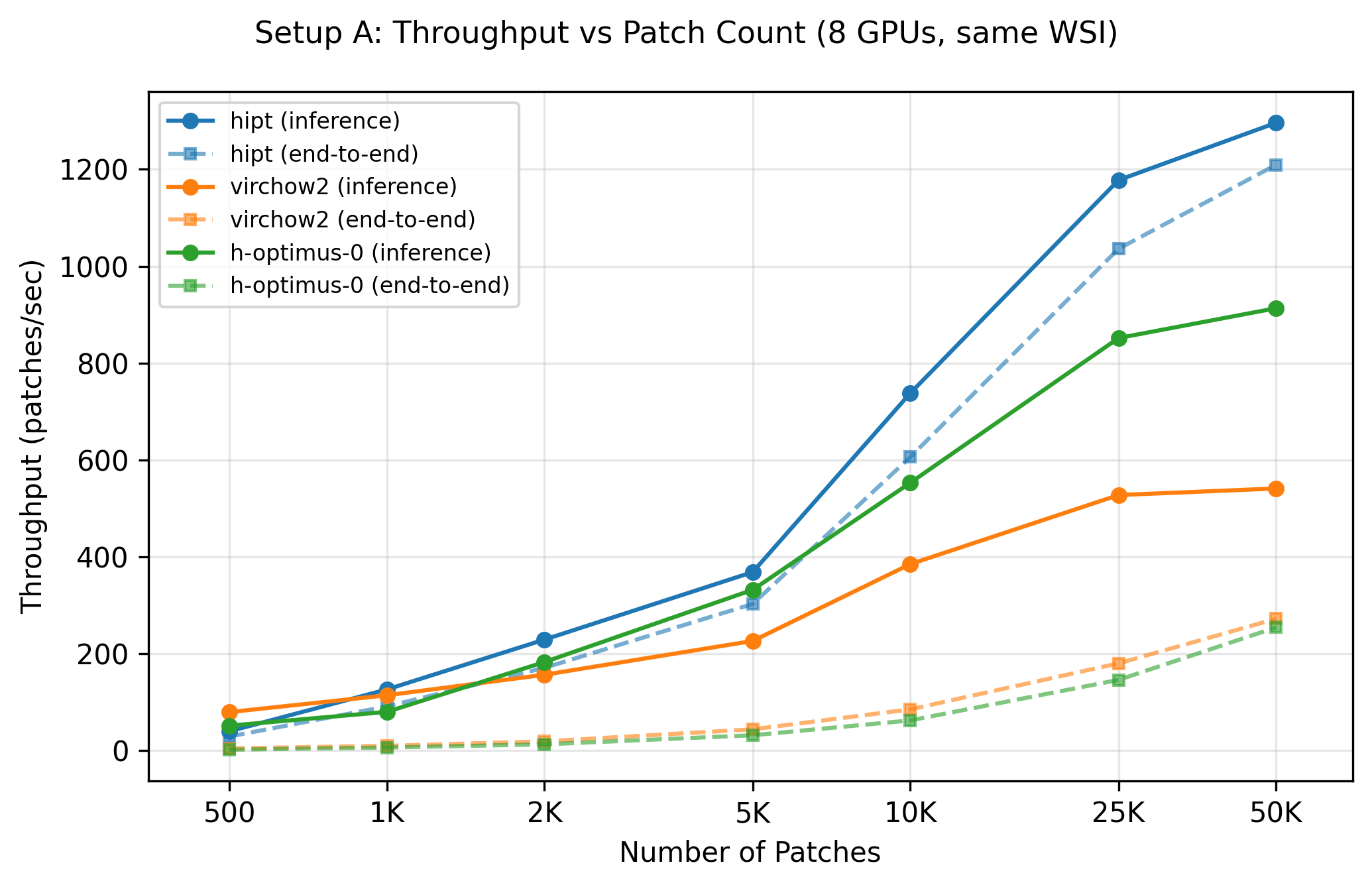}
        \caption{Throughput vs patch count.}
        \label{fig:setupA_core_throughput}
    \end{subfigure}
    \begin{subfigure}[t]{0.85\textwidth}
        \centering
        \includegraphics[width=\linewidth]{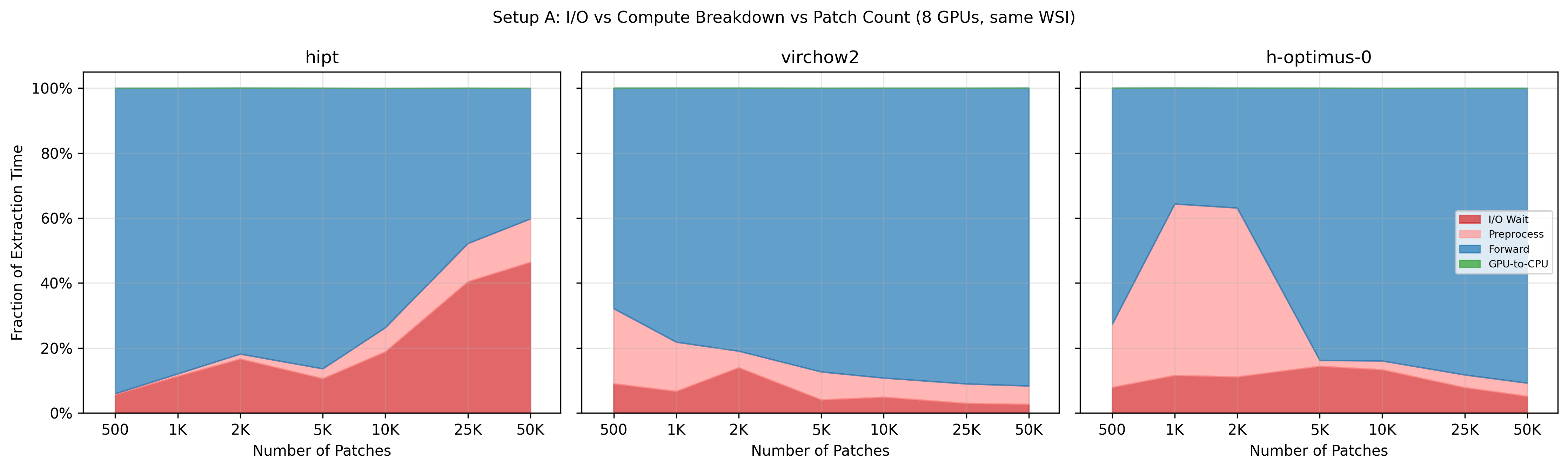}
        \caption{Extraction-time breakdown by phase.}
        \label{fig:setupA_core_breakdown}
    \end{subfigure}
    \caption{
    Setup A (Patch Sweep). As patch count increases, throughput plateaus while I/O wait grows and GPU utilization remains modest, indicating a transition from compute-bound to I/O-dominated (storage-limited) execution. This shows that increasing workload size does not improve accelerator utilization when data delivery becomes the bottleneck.
    }
    \label{fig:setupA_core}
\end{figure*}

Figure~\ref{fig:setupA_core} evaluates performance as patch count increases for a fixed WSI and hardware configuration. At small scales, throughput increases with workload size and execution remains compute-dominated. As patch count increases, throughput plateaus while I/O wait-time increases, indicating that the movement of the data becomes the limiting factor (Figure~\ref{fig:setupA_core_throughput},Figure~\ref{fig:setupA_core_breakdown}).

GPU utilization remains modest at all scales (Figure~\ref{fig:setupA_core_gpu}), despite sufficient computing capacity. This confirms that accelerators are underutilized due to input data latency rather than insufficient compute. As workload size increases, throughput flattens, I/O wait grows, and GPU utilization stays modest, indicating a transition to I/O-dominated execution even without communication overhead.

\textbf{Setup B: Strong Scaling — From Communication-Free to I/O-Dominated}

\begin{figure*}[t]
    \centering
    \includegraphics[width=0.98\linewidth]{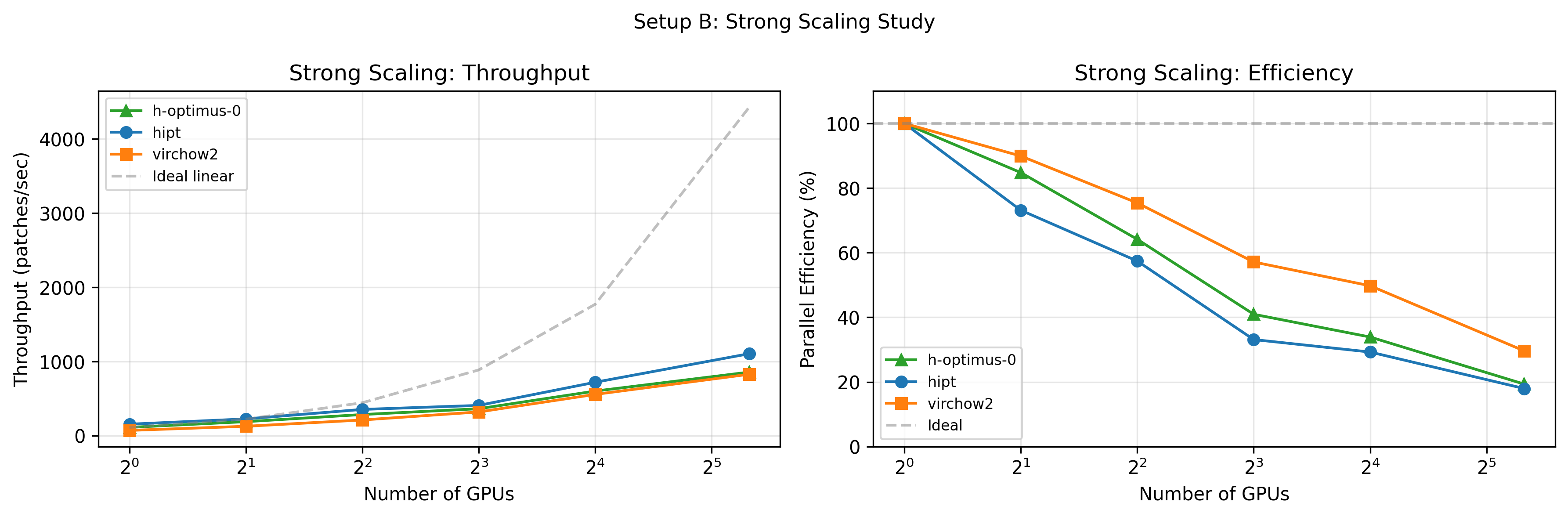}
    \caption{Setup B. Strong-scaling efficiency versus GPU count across models. Despite fully independent execution with no collective communication in the critical path, scaling deviates from linear and efficiency drops at higher concurrency. This demonstrates that performance is limited by shared storage bandwidth rather than communication overhead.}
    \label{fig:embed_setupB_scaling}
\end{figure*}

Figure~\ref{fig:embed_setupB_scaling} shows strong scaling behavior across HIPT, H-Optimus-0, and Virchow2. Throughput increases with GPU count, but scaling is not linear, with efficiency dropping below 30–50\% at higher concurrency. This degradation occurs despite fully independent execution without collective communication.

The key result is that communication-free execution (in critical path) does not guarantee scalability. Instead, performance is limited by shared storage bandwidth and metadata contention as all workers concurrently load patch data. Increasing parallelism saturates the I/O path, leading to diminishing returns and early efficiency collapse.

\begin{table}[h]
\centering
\begin{tabular}{llrrrr}
\toprule
Model & GPUs & Throughput & Speedup & Efficiency & I/O \% \\
\midrule
h-optimus-0 & 1 & 110.7 & 1.00x & 100.0\% & 1.8 \\
h-optimus-0 & 2 & 187.7 & 1.70x & 84.8\% & 2.5 \\
h-optimus-0 & 4 & 284.0 & 2.57x & 64.1\% & 3.5 \\
h-optimus-0 & 8 & 362.5 & 3.27x & 40.9\% & 6.5 \\
h-optimus-0 & 16 & 600.3 & 5.42x & 33.9\% & 10.7 \\
h-optimus-0 & 40 & 855.8 & 7.73x & 19.3\% & 15.5 \\
\midrule
hipt & 1 & 153.7 & 1.00x & 100.0\% & 35.4 \\
hipt & 2 & 224.8 & 1.46x & 73.1\% & 25.7 \\
hipt & 4 & 352.9 & 2.30x & 57.4\% & 20.3 \\
hipt & 8 & 407.1 & 2.65x & 33.1\% & 9.5 \\
hipt & 16 & 719.3 & 4.68x & 29.2\% & 10.2 \\
hipt & 40 & 1104.1 & 7.18x & 18.0\% & 15.0 \\
\midrule
virchow2 & 1 & 69.8 & 1.00x & 100.0\% & 1.2 \\
virchow2 & 2 & 125.5 & 1.80x & 89.9\% & 1.8 \\
virchow2 & 4 & 210.5 & 3.02x & 75.4\% & 2.3 \\
virchow2 & 8 & 318.9 & 4.57x & 57.1\% & 5.1 \\
virchow2 & 16 & 555.6 & 7.96x & 49.7\% & 6.6 \\
virchow2 & 40 & 826.9 & 11.85x & 29.6\% & 12.9 \\
\bottomrule
\end{tabular}
\caption{Strong scaling of embedding inference across models. Throughput increases with GPU count, but efficiency degrades and I/O fraction rises with concurrency. This indicates that scaling is limited by shared storage bandwidth rather than compute or communication.}
\label{tab:scaling}
\end{table}

\begin{figure*}
        \centering
        \begin{subfigure}[t]{0.7\textwidth}
        \centering
        \includegraphics[width=\linewidth]{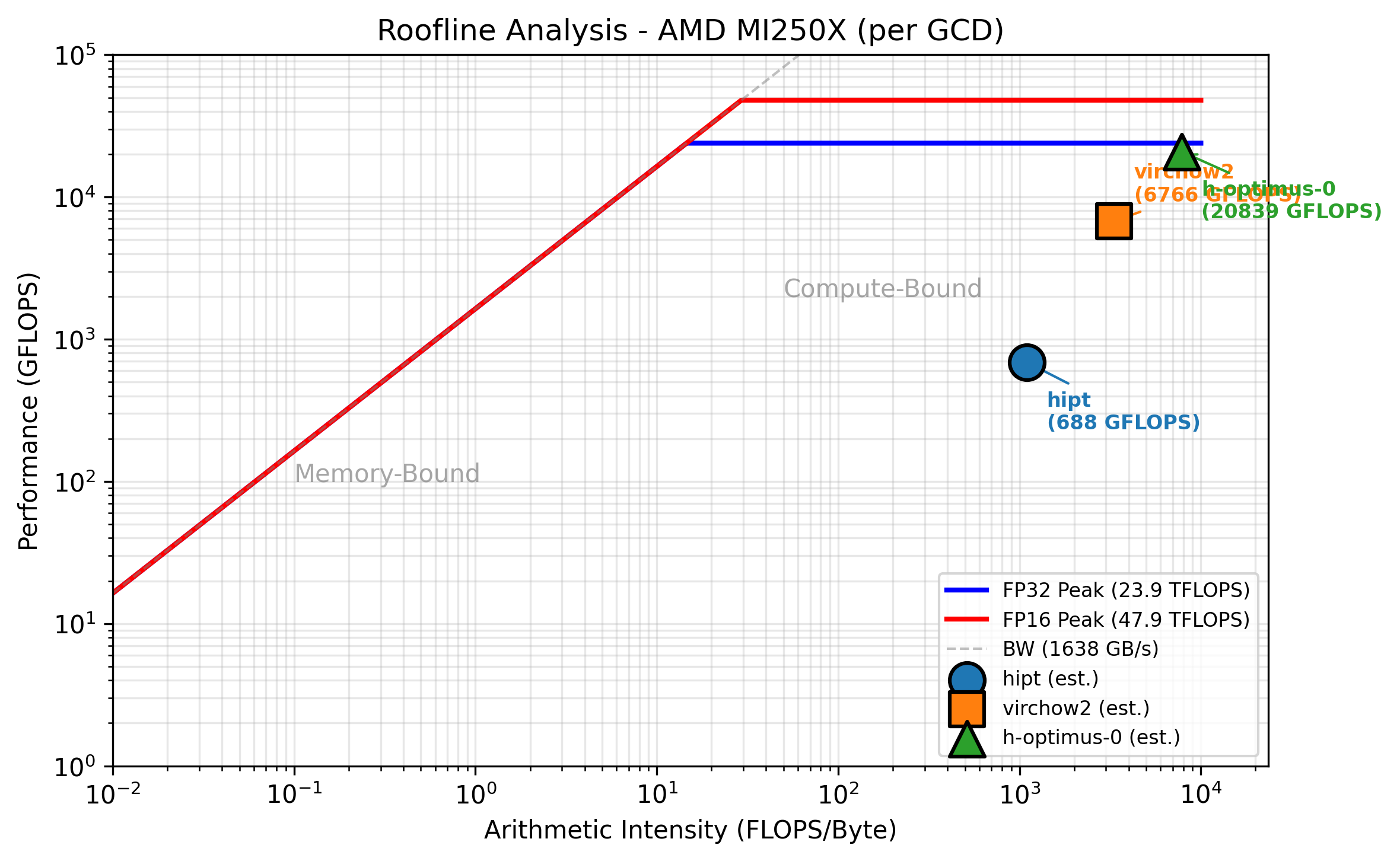}
        \caption{Roofline positioning of model kernels against hardware compute/bandwidth ceilings. We include this to distinguish kernel-local limits from system-level limits. Even when kernels operate in compute-favorable regions, the overall pipeline can remain data-movement-bound due to upstream patch delivery constraints.}
        \label{fig:embed_setupC_roofline}
    \end{subfigure}
    \begin{subfigure}[t]{0.9\textwidth}
        \centering
        \includegraphics[width=\linewidth]{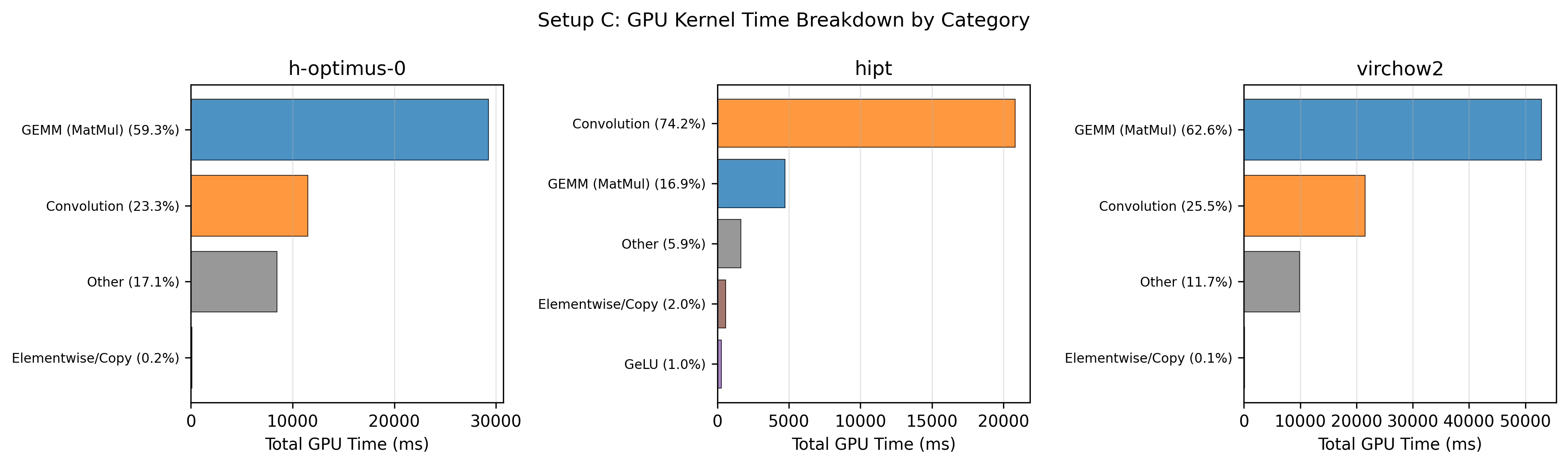}
        \caption{Kernel-category time composition (e.g., GEMM, convolution, elementwise) by model. We include this to explain model-dependent I/O sensitivity through internal compute structure: kernels may be efficient yet short enough that filesystem latency becomes exposed between batches.}
        \label{fig:embed_setupC_kernel_breakdown}
    \end{subfigure}
    \caption{Setup C (Kernel-level analysis)}
    \label{fig:SetupC}
\end{figure*}

\textbf{Setup C: Kernel-Level Analysis — Not Compute-Bound}
Kernel-level profiling (Fig.~\ref{fig:embed_setupC_roofline}) shows that model execution operates in compute-efficient regions relative to hardware limits. Across models, kernels achieve high arithmetic intensity and do not approach memory-bound ceilings, indicating that GPU execution itself is efficient.

However, kernel durations are short (Fig.~\ref{fig:embed_setupC_kernel_breakdown}), exposing latency between batches. As a result, GPUs stall waiting for data despite efficient kernel execution. This confirms that the embedding stage is not compute-bound in practice, and that performance limitations arise from upstream data delivery rather than model inefficiency.

\textbf{Setup D: Production-Scale Behavior — Orchestration and Variability}

\begin{figure*}[h]
\begin{subfigure}[t]{\textwidth}
    \centering
    \includegraphics[width=0.95\linewidth]{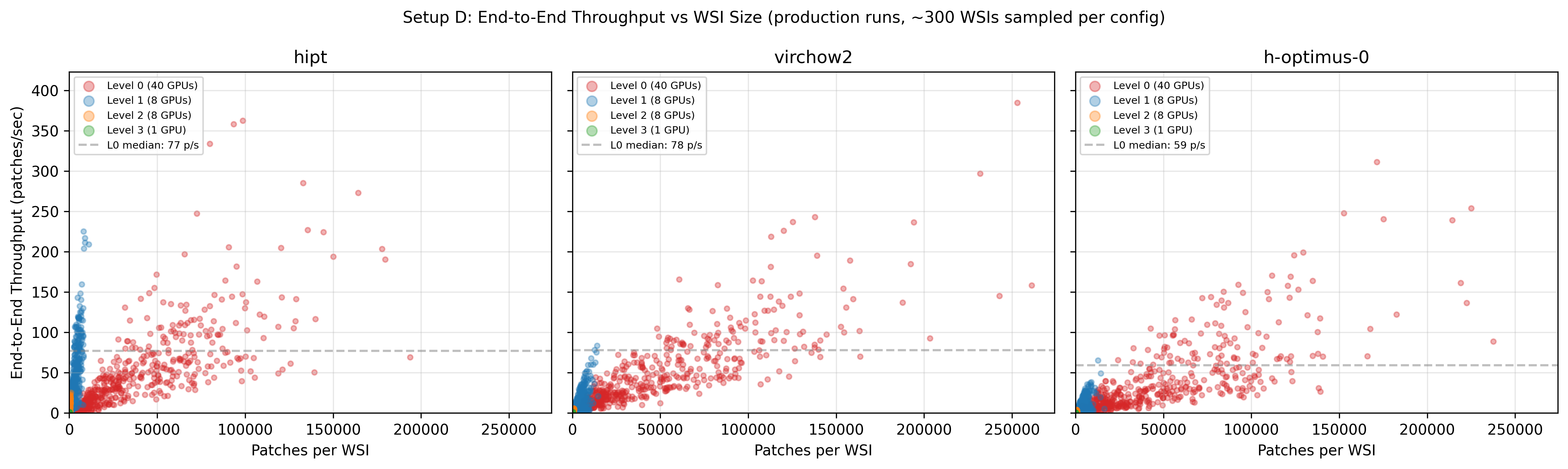}
    \caption{Setup D. Throughput versus WSI patch count across production levels. We include this to test generalization beyond controlled sweeps. Larger WSIs amortize fixed overheads better, while residual variance captures storage jitter/contention and heterogeneous slide complexity seen in real deployments.}
    \label{fig:embed_setupD_throughput_wsi}
\end{subfigure}

\begin{subfigure}[t]{\textwidth}
    \centering
    \includegraphics[width=0.95\linewidth]{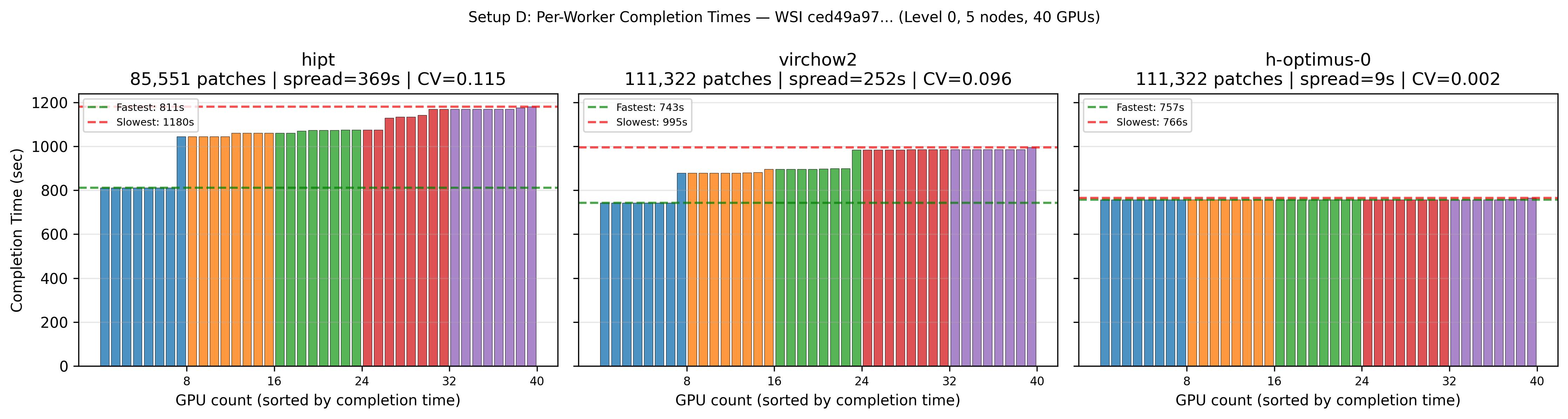}
    \caption{Setup D. Per-worker completion-time variance for representative runs. We include this because wall-clock stage time is determined by stragglers. Large spreads indicate non-trivial idle tails and motivate work-stealing and better load balancing for I/O-sensitive model settings.}
    \label{fig:embed_setupD_worker_variance}
\end{subfigure}
\caption{Setup D (Production Scale Behavior) }
    \label{fig:SetupD}
\end{figure*}

Production-scale results across 4,185 WSIs (Fig.~\ref{fig:embed_setupD_throughput_wsi}) show that throughput increases with patch count due to amortization of fixed overheads. However, significant variance remains across runs, reflecting storage contention and heterogeneous slide characteristics. Larger WSIs benefit from better amortization, while smaller WSIs suffer from disproportionate I/O costs.

Per-worker completion times (Fig.~\ref{fig:embed_setupD_worker_variance}) reveal substantial spread, with stragglers determining overall runtime. This variability highlights the importance of orchestration and load balancing in large-scale deployments. These results confirm that I/O-dominated behavior persists at production scale and is amplified by workload heterogeneity and system-level contention.

\subsection{Vector Database Ingestion Analysis}

Table~\ref{tab:hipt_scaling} shows strong scaling for VDB ingestion. Throughput increases from 22K to 188K rows/s (8.5× speedup), but efficiency peaks at intermediate scale (86.7\% at 4 nodes) and declines at higher concurrency. This non-monotonic behavior reflects the balance between parallelism and filesystem contention.

At low concurrency, available bandwidth is underutilized; at high concurrency, metadata and read contention dominate. Importantly, VDB ingestion remains I/O-bound, with performance determined by read bandwidth and write amplification during indexing rather than compute. This mirrors behavior observed in earlier stages.

Notably, the ingestion stage is embarrassingly parallel from a compute perspective: each rank independently processes a subset of embedding files and writes to a separate shard without coordination. However, all ranks share the underlying storage system, making I/O the dominant bottleneck. As concurrency increases, contention for read bandwidth and metadata access limits scalability, leading to diminishing returns.

Memory usage remains stable across configurations (approximately 0.6 GB per rank), confirming that the ingestion process follows a streaming design without significant in-memory accumulation. This indicates that performance limitations arise from external I/O constraints rather than internal resource pressure.

Overall, these results demonstrate that vector database ingestion, like earlier pipeline stages, is fundamentally I/O-bound. While shard-level parallelism enables substantial throughput gains, scalability is ultimately constrained by shared filesystem behavior. The 4-node configuration represents a practical operating point, balancing parallelism with manageable storage contention.

\begin{table*}[t]
\centering
\small
\begin{tabular}{lrrrrrrr}
\toprule
Model & Total Rows & Avg Rate & Max Rate & Files & Mean File Time (s) & Total Time (min) & RSS Max (GB) \\
\midrule
HIPT      & 6,143,506 & 2628.88 & 2775 & 14,499 & 4.27   & 1032.18  & 0.62 \\
Optimus   & 2,298,630 & 258.71  & 321  & 4,052  & 55.59  & 3754.24  & 1.68 \\
Virchow2  & 5,413,012 & 131.86  & 155  & 11,726 & 116.52 & 22772.80 & 2.74 \\
\bottomrule
\end{tabular}
\vspace{1mm}
\caption{
Performance comparison of large-scale VDB ingestion of embedding across three foundation models.
Rows are sorted by average ingestion throughput (rows/sec), highlighting substantial performance differences across embedding backbones.
HIPT achieves the highest throughput and lowest per-file processing time, while Virchow2 exhibits significantly slower ingestion despite processing a comparable number of rows.
We also observe increased memory footprint and total runtime for larger embedding sizes (e.g., Virchow2), emphasizing the importance of model efficiency in distributed vector database construction.
}
\label{tab:ingestion_performance}
\end{table*}

\begin{table*}
\centering
\small
\begin{tabular}{lrrrrr}
\toprule
Nodes & Agg.\ Rate (rows/s) & Wall Time (min) & Speedup & Efficiency (\%) & RSS Max (GB) \\
\midrule
1  & 22,216$^{*}$  & 127.9$^{*}$ & 1.00$\times$ & 100.0 & 0.58 \\
2  & 40,694        & 83.84       & 1.53$\times$ & 76.3  & 0.59 \\
4  & 86,099        & 36.90       & 3.47$\times$ & 86.7  & 0.60 \\
8  & 188,020       & 20.79       & 6.15$\times$ & 76.9  & 0.59 \\
\bottomrule
\end{tabular}
\vspace{1mm}
\caption{
\textbf{Strong scaling ablation for HIPT embedding VDB ingestion on Frontier.}
Each configuration ingests the same ${\sim}170$M embedding rows into Milvus; we vary the number of nodes (16 ranks each) and report aggregate throughput, wall-clock time, speedup, and parallel efficiency (speedup$/N$).
$^{*}$Extrapolated from observed rate; the 1-node run did not complete within the 2-hour allocation.
The 4-node configuration achieves the highest parallel efficiency ($86.7\%$) and corresponds to the production run in Table~\ref{tab:ingestion_performance}.
}
\label{tab:hipt_scaling}
\end{table*}

\section{Discussion}

Our results consistently show that large-scale WSI embedding extraction deviates from classical compute-bound scaling assumptions and instead exhibits I/O-dominated behavior across all pipeline stages. While each stage (patch generation, embedding inference, and vector database ingestion) exhibits ideal parallel complexity in isolation, end-to-end performance is governed by data movement and storage system constraints.

From a scaling perspective, patch generation nominally follows $\mathcal{O}(N/R)$ work distribution across ranks, but in practice incurs significant per-patch overhead due to small-file creation and metadata operations. As shown in Tables~\ref{tab:patch_generation_main},~\ref{tab:patch_scaling}, increasing parallelism yields sublinear speedup despite the absence of communication, indicating that filesystem contention—not computation—limits scalability. This behavior is driven by the expansion of WSIs into millions of independent patches, where metadata overhead becomes comparable to data movement.

A similar pattern emerges in the embedding stage. Although model inference is computationally efficient—as confirmed by kernel-level analysis (Fig~\ref{fig:SetupC}); GPU utilization remains modest and throughput plateaus as patch counts increase (Fig~\ref{fig:setupA_core}). The limiting factor is not compute capacity, but input data latency: GPUs stall between batches waiting for patch delivery. Strong-scaling results (Fig.~\ref{fig:embed_setupB_scaling}) further reinforce this point—performance degrades with increasing GPU count despite fully independent execution, demonstrating that communication-free designs do not eliminate scaling bottlenecks when storage bandwidth is shared.

Vector database ingestion exhibits the same fundamental constraint. While shard-parallel insertion enables throughput improvements (Table~\ref{tab:hipt_scaling}), efficiency peaks at intermediate scale and declines under higher concurrency due to read contention and write amplification. This confirms that even downstream stages, which are conceptually independent, remain bounded by storage system behavior.

These observations can be summarized by a simple performance model:
\[
T(R) = \max\left(\frac{N}{R}, \frac{N \cdot B}{\text{BW}_{\text{storage}}}\right)
\]
where \(N\) is the total number of patches to process, \(R\) is the effective compute processing rate, \(B\) is the average number of bytes transferred per patch, and \(\text{BW}_{\text{storage}}\) is the effective storage bandwidth. Increasing compute capacity reduces the compute term, \(N/R\), but does not reduce the data-transfer term, \(NB/\text{BW}_{\text{storage}}\). 
This explains the observed plateau in throughput and collapse in parallel efficiency beyond moderate scale.

Importantly, this I/O-dominated behavior is not merely a systems artifact but a direct consequence of the biomedical workload structure. WSIs are inherently high-resolution and large-scale, and meaningful analysis requires preserving patch-level granularity. This transforms compact slide data into massive collections of small, independently processed objects, making data movement unavoidable. In addition, embeddings must remain linked to their originating patches and spatial metadata to support interpretability and downstream clinical use. As a result, eliminating data movement is neither feasible nor desirable.

Instead, our results suggest a shift in system design priorities: optimizing compute alone is insufficient. Scalable WSI pipelines must explicitly account for data orchestration, storage access patterns, and intermediate data representation. The decoupled pipeline design proposed in this work addresses this requirement by isolating I/O-heavy, compute-heavy, and write-heavy stages, enabling independent optimization and reducing cross-stage interference.

Overall, our findings reposition large-scale WSI embedding extraction as a data-centric systems problem, where performance is governed not by accelerator throughput or communication, but by the ability to efficiently generate, move, and persist massive intermediate datasets. \balance

\section{Conclusion}
We presented a decoupled, I/O-aware pipeline for large-scale whole-slide image embedding extraction that separates patch generation and staging, embarrassingly parallel inference, and sharded vector database ingestion. Across controlled benchmarks and production-scale runs, we showed that this workload is fundamentally constrained by data movement and orchestration, rather than by collective communication or peak accelerator throughput. Throughput plateaus, modest GPU utilization, rising I/O fractions under scaling, and substantial VDB ingestion costs all point to the same conclusion: large-scale WSI embedding extraction is best understood as a data-centric systems problem.

Finally, the pipeline produces a persistent vector database in which embeddings are stored together with their associated metadata, including spatial and slide-level context. This representation allows downstream tasks to operate directly on embeddings without repeatedly accessing raw WSIs, avoiding redundant patch extraction and reducing I/O overhead. As a result, embedding extraction becomes a one-time data generation stage whose outputs can be reused across analyses. This is particularly important at scale, where data movement dominates runtime and repeated access to raw images is costly.


\section{Acknowledgments}
OLCF: This research used resources of the Oak Ridge Leadership Computing Facility, which is a US Department of Energy (DOE) Office of Science User Facility supported under Contract DE-AC05-00OR22725. RADIANCE: This work has been supported by DOE’s Office of Science, Office of Advanced Scientific Computing under award number DE-SC-KJ0403010. UT Battelle: This research has been authored by UT-Battelle LLC under contract number DE-AC05-00OR22725 with the DOE. 


\bibliographystyle{IEEEtran}
\bibliography{references}

\end{document}